\documentclass[conference]{IEEEtran}
\IEEEoverridecommandlockouts

\usepackage{cite}

\usepackage{amsmath,amssymb,amsfonts}
\usepackage{algorithmic}
\usepackage{graphicx}
\usepackage{hyperref}
\usepackage{textcomp}
\usepackage{xcolor}
\usepackage{soul}
\def\BibTeX{{\rm B\kern-.05em{\sc i\kern-.025em b}\kern-.08em
    T\kern-.1667em\lower.7ex\hbox{E}\kern-.125emX}}
\begin{document}

\title{\textbf{Multi-Class Neurological Disorder Prediction with Tensor Network Feature Engineering}}

\newcommand{\linebreakand}{%
  \end{tabular}\par
  \vspace{1.5em}
  \begin{tabular}{ccc}
}

\author{
Keshav Balakrishna$^{1}$, Aaryan Chityala$^{2}$, Vivan Kanna$^{3}$, Ishan Pathak$^{4}$, Harshit Ravula$^{3}$, Aaron Lee$^{2}$,\\
Alessandro Hammond$^{4}$, Moemal Al-Wishah$^{3}$, Leo Anthony Celi$^{5*}$\\[0.5em]

$^{1}$ The University of Texas at Austin, Austin, TX, USA\\
$^{2}$ University of Houston, Houston, TX, USA\\
$^{3}$ Jordan School District, Jordan, TX, USA\\
$^{4}$ Harvard University, Cambridge, MA, USA\\
$^{5}$ Massachusetts Institute of Technology, Cambridge, MA, USA
}
\maketitle

\noindent\textbf{*Corresponding Author}\\ Leo Anthony Celi\\ Massachusetts Institute of Technology\\ Cambridge, MA, USA\\ lceli@mit.edu\\

\begin{abstract}
Accurate diagnosis of neurological disorders is contingent upon advanced imaging modalities such as Magnetic Resonance Imaging (MRI), which commonly utilize sparse imaging techniques to reconstruct images from limited data, thus reducing storage and acquisition time. However, challenges remain in managing noise and preserving critical diagnostic features for effective analysis. In this study, an ensemble classifier is enriched with PARAFAC CP tensor decompositions, drawing mathematical inspiration from quantum neural network architectures but implemented entirely classically. The model was evaluated on a large, balanced clinical dataset comprising 55,160 images across 8 diagnostic categories, employing both higher and lower PARAFAC rank configurations.
Evaluated through 5-fold nested stratified cross-validation, both configurations achieved strong validation performance, demonstrating robustness to tensor network expressivity. Additionally, the proposed model achieved competitive performance relative to recent classical approaches, further underscoring the potential of quantum-inspired classical frameworks to enhance medical image analysis and support reliable clinical diagnosis. Future work will explore the integration of advanced encoding schemes, deployment on real quantum hardware, and the use of more diverse neurological datasets.
\end{abstract}


\section{Introduction}
Neurological disorders are a category of medical conditions that affect the nervous system, particularly within the spinal cord. As of 2021, an estimated 3 billion people are currently affected by neurological conditions, accounting for roughly one-third of the global population [1] [5]. Furthermore, the number of cases of brain-related disorders is expected to rise significantly to over 4.9 billion by 2050, representing an increase of over 22 percent from recent figures [4]. Neurological disorders encompass a wide array of prevalent conditions ranging from Alzheimer’s disease to cerebral aneurysms, where abnormal structural, biochemical, and electrical signals lead to serious and life-threatening symptoms such as paralysis, muscle weakness, seizures, poor coordination, and altered consciousness levels. Current diagnostic tools utilized for the monitoring of neurological-related conditions primarily utilize either X-ray or MRI scanning, generated through a process known as Sparse Imaging, where small subsets of data points construct an image [2] [3]. Thus, this illustrates the notion that information within a certain image can be represented in a limited number of significant features, consequently leading to a reduction in storage needs [3]. Although sparse imaging has various applications ranging from medical to satellite imaging, most scientific images are prone to noise artifacts outside of the original scene [10], reducing diagnostic quality from factors including poor lighting and motion during acquisition, and leading to the challenge of denoising in a manner such that crucial features, including anatomical details and texture, remain preserved without the introduction of artificial structures.
Quantum computing possesses the possibility of allowing for possible solutions to these problems, with Quantum Tensor Networks (QTNs) in particular emerging as a transformative framework to address the limitations associated with Sparse Imaging. Rooted in Quantum Many-Body Physics, QTNs can efficiently approximate high-dimensional sequences of data through the use of low-rank tensor decompositions, thus allowing for robust feature extractions from noisy MRI imaging [6]. While classical Convolutional Neural Networks (CNNs) require extensive training and updates in parameters, QTNs leverage the use of entanglement-inspired architectures, effectively modeling local and global image correlations with significantly fewer parameters [8][9]. Furthermore, recent advancements illustrate the efficacy of QTNs in medical imaging. For instance, Matrix Product State (MPS)-based models consistently outperform classical CNNs in COVID-19 CT Classification, and 3D Quantum-Inspired Self-Supervised Networks (3D-QNets) achieved volumetric segmentation without labeled data [6]. By encoding images into Quantum-State representations, these systems effectively preserve edge sharpness and texture fidelity critical for neurological diagnostics [6].
\textbf{This paper introduces a solution using PARAFAC tensor decompositions for optimal MRI image classification, enhancing denoising and preserve features. By addressing the challenges of anatomical fidelity and noise reduction, the proposed model advances the development of classical solutions with quantum-inspired mathematical structures for the early detection of neurological disorders.}

\section{Literature Review}
\subsection{The Foundation of Quantum Tensor Networks}
Tensor networks$-$initially used for quantum many-body physics$-$now provide an advanced mathematical framework for machine learning applications. Survey studies have shown that tensor networks provide structured approaches to decompose high-dimensional tensors via contraction of lower-rank tensors, illustrating their effectiveness for quantum systems with very complex organization [11]. In context, tensor networks involve efficiently encoding quantum states, which allows computation technology to manipulate large spaces of quantum data. The most commonplace tensor network architectures include, but are not limited to: Matrix Product States, Projected Entangled Pair States (PEPS), Tree Tensor Networks (TTN), and Multi-scale Entanglement Renormalization Ansatz (MERA). Rieser et al. (2023) outlined how the new area of quantum machine learning is bringing these designs, which were initially created for quantum theory, back to the quantum domain after proving to be a viable machine learning paradigm [11]. Depending on the application domain, each design provides a unique method for controlling entanglement qualities while preserving computing efficiency, with varying benefits. The team concludes that the construction of tensor networks (TNs), inspired by quantum principles, facilitates a seamless translation of the concept into quantum computations. Tensor nodes are implemented through multi-qubit gates, with the incoming and outgoing qubits representing the bonds of the node [11]. The use of these networks represents a pivotal development for applications in quantum computing, where complex systems can be both modeled and computed by leveraging the abilities of these networks to reduce computational requirements. This directly corresponding behavior enables scientists and engineers to gain insights into classical and quantum implementations in solving machine learning problems, thus facilitating the development of hybrid classical-quantum algorithms with more potential than either of the two standalone approaches.

\subsection{Using Quantum Tensor Networks for Medical Imaging Analysis in Neurological Disorders}
The utilization of tensor networks in medical imaging has demonstrated significant advantages over conventional approaches, with recent quantum computing research focused on image denoising and segmentation. One such study involved brain MR images obtained voluntarily from five individuals, who were treated with varying levels of artificial noise. The modified images were then denoised via a shrinkage convolutional neural network (SCNN) and dDLR [15]. Note: In the context of medical imaging, particularly MRI, dDLR stands for Deep Learning-based Reconstruction. Kidoh et al. (2019) concluded that dDLR reduces image noise while preserving image quality on brain MRI images, illustrating the potential for hybrid models, involving image denoising methods incorporated with quantum machine learning, to process large and complex datasets in a more efficient manner, potentially leading to faster and more accurate diagnoses of neurological disorders [15]. Recently, advancements in QITNs have opened new avenues for medical imaging analysis, particularly in the context of neurological disorders. For instance, Konar et al. (2023) introduced a three-dimensional self-supervised TN for the volumetric segmentation of medical images. The study was conducted through a quantum-inspired and self-supervised Voxel-Wise Neural Network referred to as 3D-QNet, a model that is relevant to volumetric medical image segmentation. The self-supervised learning was proposed via a tensor representation of weight vectors for multidimensional image data employed in the suggested 3D-QNet for the task [6]. This is highly relevant for brain imaging, where volumetric analysis is crucial for detecting and classifying neurological disorders. The experimental results indicated that the proposed 3D-QNet performs optimally for complete brain tumor segmentation for several volumetric MRI images. By implementing self-supervised objectives, the model can learn robust feature representations from raw brain MRI images, making it suitable for sparse data scenarios often encountered in research involving neurological disorders [6]. In the context of sparse image classification, QITNs can capture complex spatial dependencies in brain MRI data, potentially improving the detection of subtle patterns associated with neurological disorders such as Alzheimer’s disease, stroke, or multiple sclerosis [6] [13].

\subsection{Future Directions and Potential Challenges}
QITNs have demonstrated a significant potential in sparse imaging for neurological disorder classification, offering advantages and superior results in both algorithm accuracy over classical approaches. Shahriyar and Tanbhir (2025) suggest integrating quantum-inspired TNs with classical neural network layers, such as in hybrid quantum-classical models, to further enhance feature extraction and classification performance for increasingly complex brain MRI datasets [13]. In addition, advances in quantum technologies and AI-induced processing can potentially enable the use of portable and real-time devices for neurological disorder diagnostics. Particularly, the fusion of QITNs with quantum sensors can expand access to sophisticated diagnostics beyond traditional clinical workflows [12]. Future research involving QITNs will likely involve their optimization via the design of tensor decompositions and activation functions to increase convergence rates and segmentation accuracy in volumetric MRI analysis [6]. As medical imaging datasets become larger and more complex over time, QITN models capable of correct training and inference on big data will become more necessary. This includes exploring novel TN structures and quantum-inspired heuristic algorithms for optimizing exponentially large parameters [14].
However, numerous challenges must be addressed before a widespread adoption and practical deployment of QITNs in brain MRI classification for neurological disorders. Firstly, these models are designed for classical hardware, which may be prone to erroneous connection and qubit noise. Additionally, QITN-based models must be rigorously validated against pre-existing diagnostic standards, and transparent decision-making is needed to build trust among clinicians and regulatory bodies. It is also important to note that although QITNs can reduce the number of parameters compared to deep neural networks, training and optimizing these models—especially in higher-dimensional settings—can still be computationally demanding. Developing efficient training algorithms and leveraging parallel programming will be pivotal to overcoming these barriers [6].

\section{Methodology}
\subsection{Dataset Description and Class Extraction}
\textbf{All experiments were conducted on a single NVIDIA T4 GPU (15 GB VRAM) provided by Google Colaboratory, with 12.7 GB system RAM. PARAFAC 
feature extraction was performed in parallel across all available CPU cores using \texttt{joblib}, requiring approximately 10-15 minutes 
for the full dataset of 16,390 images. Total nested cross-validation training time was approximately 100 minutes, with a mean of 20 minutes 
per outer fold. In this study, a Hybrid Ensemble Classifier using Random Forest, in combination with TN features generated through Parallel Factor Analysis (PARAFAC) (CP) tensor decompositions}, was utilized. The model was trained on a total of $55,160$ human brain MRI images, associated with three neurological disorders (Alzheimer's Disease, Brain Tumors, and Multiple Sclerosis), sourced from the Multi-Class Neurological Disorders (MCND) dataset following oversampling to create a balanced class distribution. The model was trained to classify $8$ classes in the MCND dataset labeled $0-7$, with $6,895$ images per class. Additionally, for each validation fold, $44,128$ images were used for training, while $11,032$ images were used for testing. Each folder was then mapped into a class label, forming the dictionary:
\begin{equation}\label{eq:1}
    C = \{c_1, c_2, \dots , c_K\}
\end{equation}
for a total of $K=8$ distinct classes. Each image file is associated with a class ID based on the folder name. Mathematically, let $N_0$ denote the initial size, i.e. the number of raw images, calculated as the sum of files found across all folders. This metric can be determined by applying the following equation:
\begin{equation}
    N_0 = \sum_{k=1}^K n_k^{(0)}
\end{equation}
where $n_k^{(0)}$ is the raw image count in class $c_k$ prior to data cleaning.

Following rigorous pre-processing steps, the dataset was reduced to a cleaned and distinct set of $N_1 = 16,390$ images. In order to resolve inherent class imbalance in the dataset, the Random OverSampling (ROS) method was applied exclusively withing each training fold after the stratified train/test split, replicating images in minority classes to equalize all classes at $6,895$ samples each. This ensures that oversampled duplicates strictly within the training partition and cannot appear in the held-out test out, which prevents data leakage. Class weights were calculated on the original pre-oversampling distribution to maintain meaningful loss weighting after resampling, yielding a final balanced dataset size of
$$N = K \times n_k = 8 \times 6,895 = 55,160$$
This balanced dataset ensures each category within the trained classifier, enriched with TN features, achieves equal representation, improving classification robustness across all three relevant neurological disorders.

\subsection{Image Loading and Cleaning}\
In order to ensure high-quality, consistent data across diverse brain MRI sequences, a rigorous pre-processing pipeline was implemented for all images prior to training the model. Each image was first read from its corresponding class folder using robust handling packages, where the original orientation was corrected by applying its Exchangeable Image File Format (EXIF) metadata to account for variations in scanner position or storage procedures, significantly reducing the likelihood of classification errors due to spatial misalignment. Subsequently, the images were converted to grayscale, simplifying the data by converting three color channels (RGB) to one, as color properties are non-informative in MRI modalities and can introduce noise or bias. 

To standardize the input images and ease computational efforts, each grayscale image was then resized to a fixed resolution of $64 \times 64$ pixels via bilinear interpolation. Then, rigorous quality control was performed via removal of images exhibiting corrupted data, in the form of Not a Number (NaN) or infinite pixel values, either of which can significantly disrupt numerical computations. Additionally, images that retained uniform values of pixel intensity, were discarded because they do not provide any constructive, diagnostic information. The set of cleaned images $\{x_i\}$ was then filtered to remove duplicate entries by comparing flattened pixel arrays, ensuring the non-existence of any two identical images, thus eliminating any possible bias caused by redundant data points. Eventually, the initial raw dataset of $N_0$ images was transformed to a clean set containing $N_1$ distinct, valid, and analysis-ready images. The aforementioned cleaning process is pivotal to ensure that machine learning models receive high-quality data, enhancing the classification results' robustness and generalizability.
\begin{figure}
    \centering
    \includegraphics[width=0.4\linewidth]{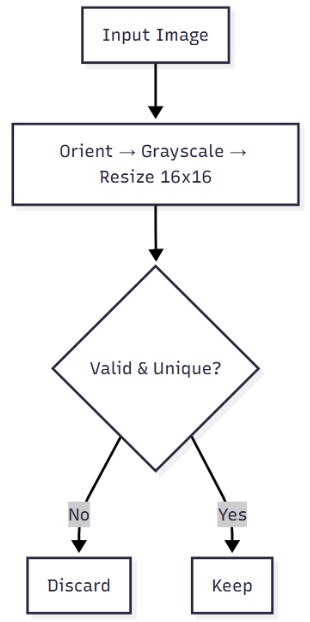}
    \caption{Diagram of streamlined image pre-processing: images are oriented, converted, resized, then checked for validity and uniqueness before inclusion in the dataset.}
    \label{fig:placeholder}
\end{figure}

\subsubsection{Corrected Orientation}
A subset of MRI images within the MCND dataset were equipped with EXIF metadata, which is responsible for storing orientation information. Specifically, it encodes how an image should experience a series of flips and rotations to appear upright and conventionally oriented when displayed within the set. However, no image-processing tool is fully effective$-$they do not apply all orientation flags, which can lead to inconsistent image display and degraded downstream analysis. To navigate this challenge, the pre-processing pipeline was equipped with an automatic orientation correction step. Thus, upon loading each image, the algorithm reads the EXIF Orientation tag$-$which may indicate one of several positions, e.g. 90$^\circ$, 180$^\circ$, or 270$^\circ$, or mirrored variants.

Specifically, the correction transformation is applied prior to processing in order to ensure homogeneity across the entire dataset. Define an image $I$ as a two-dimensional array of pixel intensity values. The EXIF Orientation tag $O(I)$ determines a transformation $T_O$ such that the corrected image $I'$ is given by the following:
\begin{equation}\label{eq:2}
    I' = T_O(I)
\end{equation}
where $T_O$ belongs to a discrete rotation and reflection operation set standardized by the EXIF metadata specification. Via $T_O$, the pre-processing pipeline guarantees every image to share a common orientation reference frame, effectively minimizing issues associated with sensor-specific or device-dependent rotations within the dataset. In the context of medical imaging, orientation correction is critical, as correct alignment prevents the analysis model from learning false rotational features and supports meaningful comparison across dataset samples, preserving the integrity of the original data.

\subsubsection{Grayscale Conversion}
To minimize the impact of confounding factors derived from color transformation, along with standardizing the original dataset, all MRI images were converted into an 8-bit, single-channel grayscale format. This transformation is fundamental because color channels typically carry no diagnostic significance and can therefore be irrelevant; additionally, the color channels may even introduce unwanted noise or other artifacts from certain storage formats. Via this conversion, the complexity of the original input data is moderately reduced, enabling more efficient and focused feature extraction and classification.

The original image initially consists of three color channels$-$red, green, and blue$-$at each pixel, denoted by $I_{RGB}(x,y) = (R, G, B)$, is converted into a single intensity value $I_{gray}(x,y)$. The mathematical computation is expressed as a weighted sum of color channels that approximates human perceptual brightness, determined by the eye's relative sensitivity to each color:
\begin{equation}\label{eq:3}
    \footnotesize{I_{gray}(x,y) = 0.2989 \cdot R(x,y) + 0.5870 \cdot G(x,y) + 0.1140 \cdot B(x,y)}
\end{equation}
The grayscale intensity values are then quantized into the 8-bit range $0-255$, where $0$ represents fully black and $255$ represents fully white. This separation ensures compatibility with standard image processing tools and maintains memory-efficient data without compromising the visibility of essential MRI image structures. Overall, conversion into this consistent 8-bit grayscale format eliminates scanner-specific color mapping variability, thus easing the representation of TN decomposition and classification tasks. In addition, the analysis model can concentrate on more significant intensity fluctuations for differentiating neurological disorder, rather than learning irrelevant color mapping features. Thus, the uniform grayscale representation supports reproducible and and robust outcomes in machine learning models.

\begin{figure}[htbp]
    \centering
    \includegraphics[width=0.2\textwidth]{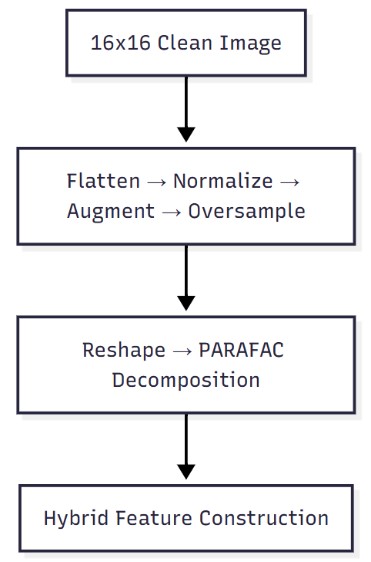} 
    \caption{Diagram of pre-processing and feature extraction pipeline. Flattening applies to the PARAFAC branch only. The CNN branch processes full 2D images.}
    \label{fig:pipeline}
\end{figure}

\subsubsection{Flattening, De-duplication, and Dataset Summary}
With correct orientation, a uniform grayscale representation, and resizing, each cleaned image was systematically prepared for the following analysis via flattening and de-duplication steps. \textbf{The resizing operation produced images of $64 \times 64$ standardized dimension pixels since the reduction in input dimensionality helps ensure TN decompositions, featured maps, and nested cross-validation remain computationally and memory feasible on classical hardware. Additionally, a very small fixed resolution simplifies FLOP accounting and makes the comparison high- vs low- rank configurations easier, at the cost of potentially discarding clinically relevant fine-grained structures.} Next, for the PARAFAC branch, each image was flattened into a one-dimensional array, mapping the two-dimensional arrangement of pixels into a single feature vector for tensor decomposition. The CNN branch (ResNet-18) receives the full 2D image directly, preserving spatial structure through its convolutional layers throughout feature extraction. Specifically, for image $I_r \in \mathbb{R}^{64 \times 64}$, the flattening operation is given by:
\begin{equation}\label{eq:1}
    x = \texttt{vec}(I_r)
\end{equation}
where the \texttt{vec} operator denotes matrix vectorization via pixel value traversing in a consistent order, usually row-major. The transformation ensures compatibility with both classical ML algorithms and tensor decomposition procedures, which require static and uniform inputs.

To build a model with high generalizability and avoidance of additional bias stemming from redundant samples, it is essential to ensure the uniqueness of training data. As a result, all flattened vectors were checked for duplication using a data frame-based approach. In particular, if two images were found to produce identical $4096$-dimensional vectors, only one copy was preserved in the dataset, while any duplicates were dropped. Deduplication was achieved via Pandas' \texttt{duplicated()} method applied to the set of all flattened image vectors, further enhancing data quality, without artificially inflating any particular imaging scheme or class.

Ultimately, the final collection of vectors served as the origin for intensity normalization, data augmentation, balancing, tensor feature extraction, and model training. The size of the curated dataset reflects the strong quality control applied, including solely those images suitable for robust, precise, and unbiased downstream learning. Overall, the comprehensive approach to image flattening and de-duplication ensures an accurate representation of real-time clinical workflows and diagnostic modeling using quantum-inspired mathematical structures.

\section{Results and Analysis}
\textbf{The performance of the proposed hybrid ensemble classifier, integrating random forests with PARAFAC tensor decompositions}, was evaluated on the MCND dataset using two principal configurations: a higher rank configuration (rank = 16) and a lower rank configuration (rank = 3) All results are reported from balanced experiments (each class: 6,895 samples, total: 55,160), including comprehensive image cleaning and augmentation. \textbf{The performance of the proposed model was evaluated through five-fold nested stratified cross-validation, with reported metrics for both inner and outer folds test sets to assess generalization in a comprehensive manner}.

\subsection{Aggregate Learning and Validation Metrics *All tables are new/recent edits in this subsection}
Tables I, II, and III summarize the cross-validation performance metrics for both rank configurations. Both configurations achieved near-perfect training performance, with slight variations in generalization metrics observed on held-out validation sets. 
\begin{table}[htbp]
\centering
\scriptsize
\caption{Fold-wise Metrics: Higher Rank Configuration (Rank = 16) - Training and Validation (Inner CV). *New edit made to paper}
\begin{tabular}{c|cccccc}
\hline
Fold & Train ACC & Train F1 & Val ACC & Val Prec & Val Rec & Val F1 \\
\hline
1 & 1.0000 & 1.0000 & 0.9492 & 0.9534 & 0.9492 & 0.9470 \\
2 & 1.0000 & 1.0000 & 0.9448 & 0.9479 & 0.9448 & 0.9422 \\
3 & 1.0000 & 1.0000 & 0.9505 & 0.9546 & 0.9505 & 0.9483 \\
4 & 1.0000 & 1.0000 & 0.9505 & 0.9545 & 0.9505 & 0.9484 \\
5 & 1.0000 & 1.0000 & 0.9512 & 0.9545 & 0.9512 & 0.9491 \\
\hline
Mean & 1.0000 & 1.0000 & 0.9493 & 0.9530 & 0.9493 & 0.9470 \\
Std & 0.0000 & 0.0000 & 0.0026 & 0.0029 & 0.0026 & 0.0028 \\
\hline
\end{tabular}
\label{tab:highent}
\end{table}
\vspace{0.5cm}
\begin{table}[htbp]
\centering
\scriptsize
\caption{Fold-wise Metrics: Lower Rank Configuration (Rank = 3) - Training and Validation (Inner CV). *New edit made to paper}
\begin{tabular}{c|cccccc}
\hline
Fold & Train ACC & Train F1 & Val ACC & Val Prec & Val Rec & Val F1 \\
\hline
1 & 0.9994 & 0.9994 & 0.9463 & 0.9502 & 0.9463 & 0.9434 \\
2 & 0.9992 & 0.9992 & 0.9445 & 0.9484 & 0.9445 & 0.9419 \\
3 & 0.9992 & 0.9992 & 0.9476 & 0.9529 & 0.9476 & 0.9446 \\
4 & 0.9992 & 0.9992 & 0.9461 & 0.9513 & 0.9461 & 0.9431 \\
5 & 0.9993 & 0.9993 & 0.9502 & 0.9541 & 0.9502 & 0.9476 \\
\hline
Mean & 0.9993 & 0.9993 & 0.9470 & 0.9514 & 0.9470 & 0.9441 \\
Std & 0.0001 & 0.0001 & 0.0021 & 0.0022 & 0.0021 & 0.0022 \\
\hline
\end{tabular}
\label{tab:lowent}
\end{table}

\vspace{0.3cm}
\begin{table}[htbp]
\centering
\scriptsize
\addtolength{\tabcolsep}{-3pt} 
\caption{Outer Fold Test Metrics for Nested CV: Higher and Lower Rank Configurations.}
\begin{tabular}{c|cccc|cccc}
\hline
Fold & \multicolumn{4}{c|}{Higher CP Rank (TN Rank=16)} & \multicolumn{4}{c}{Lower CP Rank (TN Rank=3)} \\
\hline
 & ACC & PREC & REC & F1 & ACC & PREC & REC & F1 \\
\hline
1 & 0.9478 & 0.9514 & 0.9478 & 0.9454 & 0.9464 & 0.9505 & 0.9464 & 0.9435 \\
2 & 0.9474 & 0.9510 & 0.9474 & 0.9450 & 0.9456 & 0.9501 & 0.9456 & 0.9428 \\
3 & 0.9497 & 0.9535 & 0.9497 & 0.9474 & 0.9458 & 0.9504 & 0.9458 & 0.9424 \\
4 & 0.9485 & 0.9534 & 0.9485 & 0.9464 & 0.9458 & 0.9506 & 0.9458 & 0.9429 \\
5 & 0.9514 & 0.9546 & 0.9514 &
 0.9493 & 0.9506 & 0.9541 & 0.9506 & 0.9481 \\
\hline
Mean & 0.9490 & 0.9528 & 0.9490 & 0.9467 & 0.9468 & 0.9512 & 0.9468 & 0.9439 \\
Std & 0.0016 & 0.0015 & 0.0016 & 0.0017 & 0.0021 & 0.0017 & 0.0021 & 0.0024 \\
\hline
\end{tabular}
\addtolength{\tabcolsep}{3pt} 
\end{table}

Based on the above results, both CP rank configurations achieved near-perfect learning performance during the training and validation stages. The strong similarity observed across all generalization metrics (precision, accuracy, recall, and F1) all demonstrate the stability and robustness and consistency of the hybrid modeling framework across varying tensor expressivity levels.

\subsection{Confusion Matrix Analysis and Per-Class Trends}
\begin{figure}[htbp]
    \centering
    \includegraphics[width=0.36\textwidth]{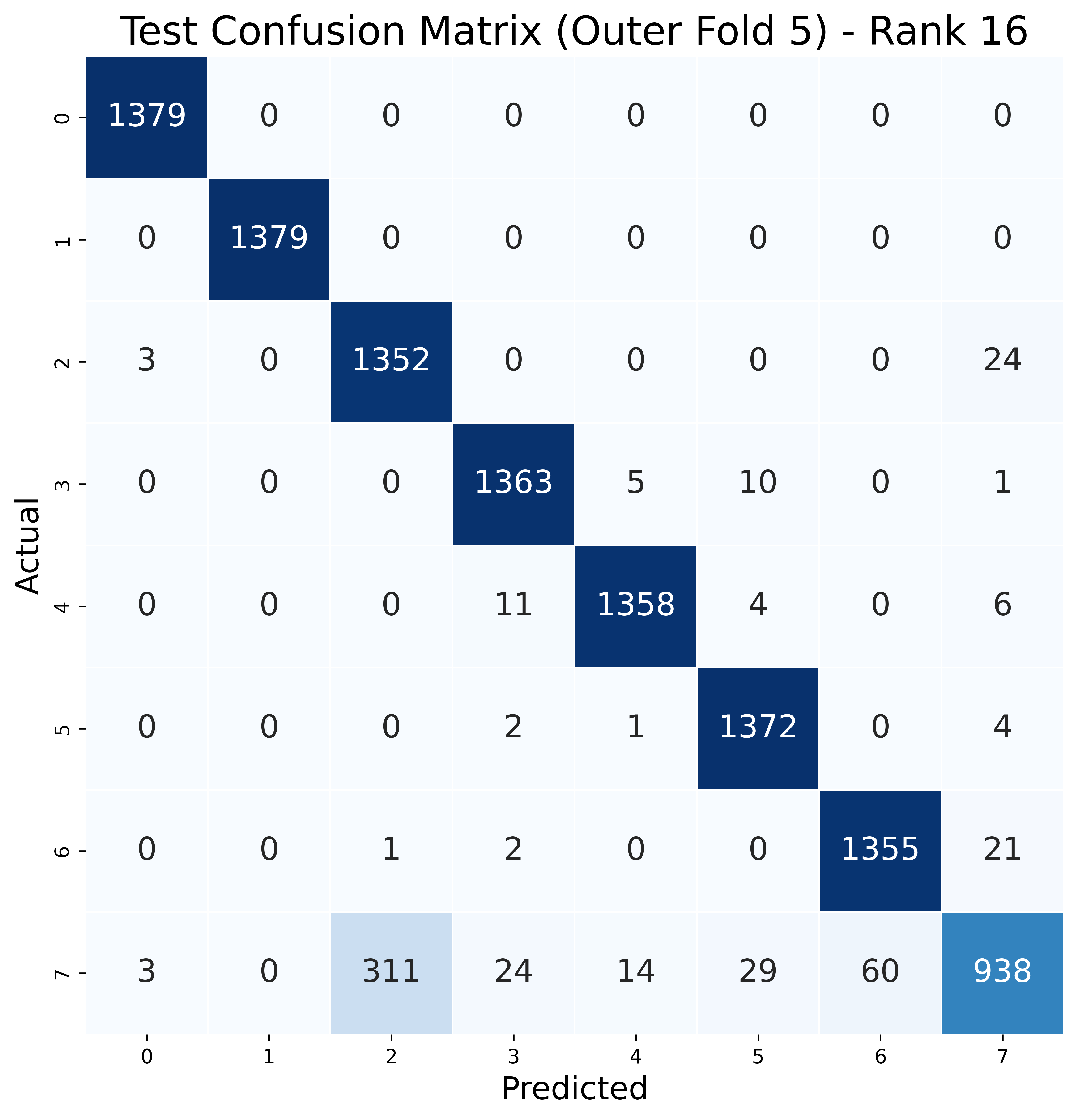} 
    \caption{Fold 5 validation CM (Rank = 16)}
    \label{fig:pipeline}
\end{figure}
\begin{figure}[htbp]
    \centering
    \includegraphics[width=0.36\textwidth]{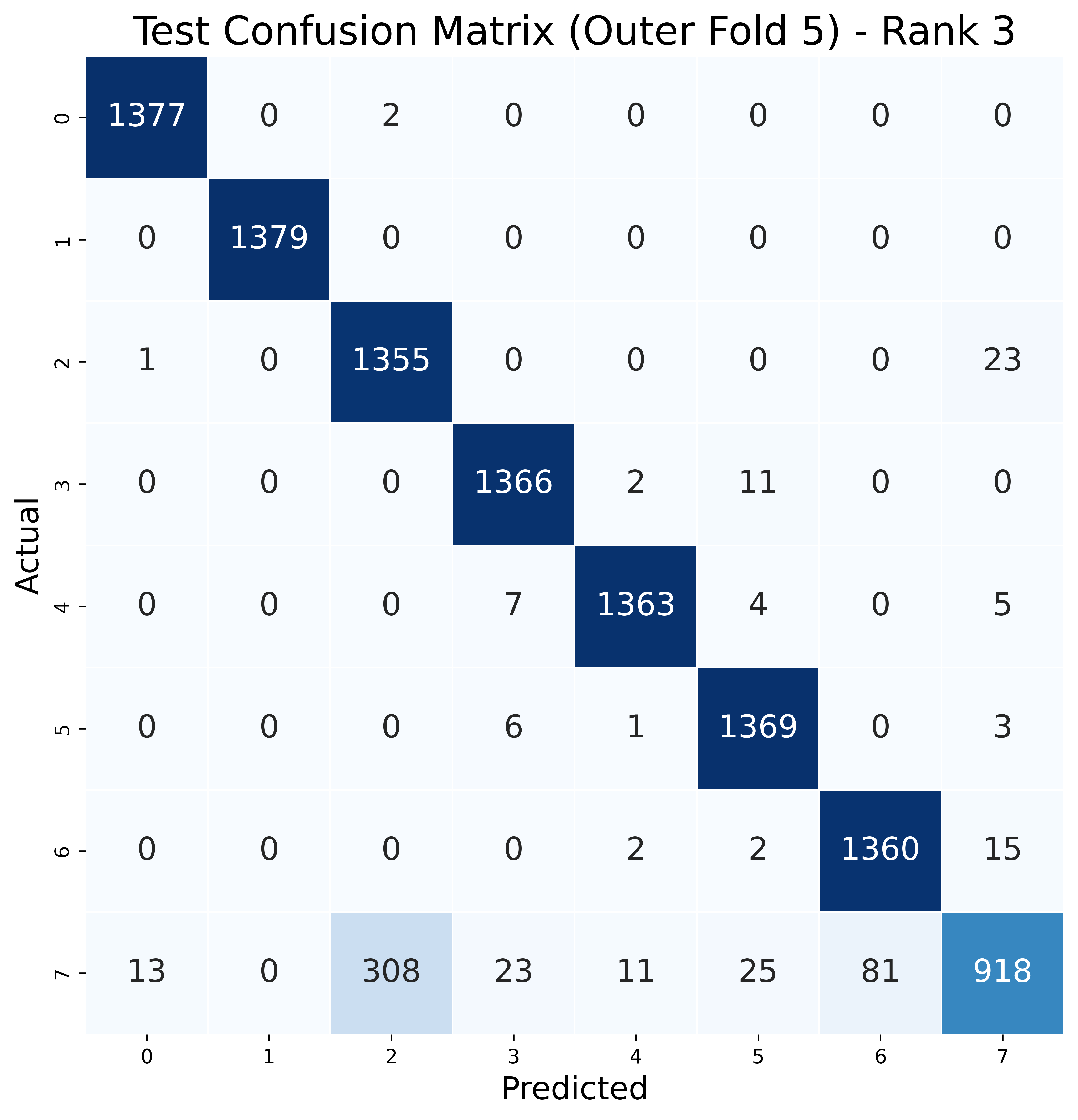} 
    \caption{Fold 5 validation CM (Rank = 3)}
    \label{fig:pipeline}
\end{figure}
\textbf{Figures 3 and 4 display the validation confusion matrices for the fifth outer fold under higher and lower rank configurations, respectively.} These results illustrate excellent discrimination among all neurological disorder classes, with mostly correct predictions concentrated along the diagonals. In both configurations, the maximum confusion occurred in class label 7 (Normal), which was occasionally misclassified as mild to moderate Alzheimer’s or other tumor categories$-$demonstrating the clinical difficulty in separating subtle normal/pathological edge cases. \textbf{In both cases, class labels 0 through 6 were identified with high accuracy, underscoring the effectiveness of the proposed framework.} The off-diagonal error trends were nearly identical for both ranks, further illustrating the reliable and data-dependent nature of observed deviations.

\textbf{Table 4 shows the FLOPs comparison of the proposed model of this study in both rank configurations and classical CNN architectures from a 2025 study. The proposed model, which had roughly 42 million FLOPs at rank 3 and 226 million at rank 16, are significantly lower than large CNN architectures such as AlexNet, yet larger than very small models such as LeNet (which uses a small input image classifier and is a Kolmogorov-Arnold CNN variant) and Tabular CNN (which has a smaller input and primarily uses tabular data). This suggests a balance in the computational efficiency for the proposed model, requiring less compute than large CNN architectures while maintaining strong predictive and classification performance competitive with such architectures.}

\begin{table}[ht]
\centering
\caption{FLOPs Comparison from [17] *This table is a new edit}
\small
\begin{tabular}{p{3.5cm} p{3cm} p{5cm}}
\hline
Model & FLOPs \\
\hline
This Study (Rank 3) & 1,813,869,832 \\
This Study (Rank 16) & 1,814,295,816\\
Tabular CNN [17] & 37,853,010 \\
LeNet (standard CNN) & 429,128 \\
LeNet KAN & 3,298,728 \\
AlexNet (standard CNN) & 714,197,696 \\
AlexNet KAN & 1,611,568,352 \\
\hline
\end{tabular}
\vspace{0.3em}
\begin{minipage}{\columnwidth}
\footnotesize\textit{Note: The proposed model uses 64$\times$64 input for the PARAFAC branch and 224$\times$224 for the CNN branch. Baseline models from the literature were trained and evaluated at their native resolutions. Direct FLOPs and accuracy comparisons should be interpreted with this difference in mind.}
\end{minipage}
\label{tab:flops_benchmark}
\end{table}

Table 5 reports validation accuracy from both higher and lower rank states of the current study, alongside metrics from leading pre-trained classical convolutional neural networks, as reported in recent literature. The classifier, in both its higher and lower rank settings, delivers a strong performance, reaching roughly a 94.90\% (higher CP rank) and 94.68\% (lower CP rank) mean validation accuracy for the outer folds. By comparison, VGG-19 achieved 99.48\%, VGG-16 99.00\%, and ResNet50 97.92\%, while Inception V3 reached 81.25\% validation accuracy. Based on the results, while large classical CNN models currently set the highest accuracy benchmarks for the medical neuro-imaging domain, the proposed model demonstrates very strong and reproducible performance that is highly competitive and valuable for clinical diagnostic pipelines, \textbf{while also requiring significantly less compute when compared to large CNN architectures.}
\newline
\begin{table}[htbp]
\centering
\scriptsize
\caption{Benchmark Validation Accuracy Comparison with Classical CNNs.}
\begin{tabular}{lcc}
\hline
Model & Val. Acc. (\%) & Source \\
\hline
Our Approach (Higher CP Rank) & 94.90 & This Study \\
Our Approach (Low CP Rank)  & 94.68 & This Study \\
VGG-19 (Pre-trained)          & 99.48 & [16] \\
VGG-16 (Pre-trained)          & 99.00 & [16] \\
ResNet50 (Pre-trained)        & 97.92 & [16] \\
Inception V3 (Pre-trained)    & 81.25 & [16] \\
\hline
\end{tabular}
\vspace{0.3em}
\begin{minipage}{\columnwidth}
\footnotesize\textit{Note: The proposed model uses 64$\times$64 input for the PARAFAC branch and 224$\times$224 for the CNN branch. Baseline models from the literature were trained and evaluated at their native resolutions. Direct FLOPs and accuracy comparisons should be interpreted with this difference in mind.}
\end{minipage}
\label{tab:flops_benchmark}
\end{table}
The results from Table 3 suggest that validation accuracy remains nearly identical across high and low-rank models, indicating that greater tensor network rank does not substantially improve generalization. The small observed gains likely reflect added expressivity but lie within statistical noise. Thus, the framework achieves robust, efficient performance without relying on high rank, making it well-suited for large-scale medical imaging. The model, therefore, demonstrates strong potential for large-scale, multi-class brain MRI classification, with both configurations yielding stable and robust results across folds. Most errors occur in clinically nuanced cases that remain difficult even for human experts and classical models. \textbf{Overall, the method is competitive, reproducible, and offers a quantum-inspired, yet classical, alternative for medical image analysis.}

\subsection{Branch Ablation}

To determine the individual contribution of each model component, a branch ablation study was conducted comparing three variants: CNN-only, PARAFAC-only, and the full fused model. Each variant was evaluated using 3-fold stratified cross-validation with macro F1 as the primary metric. Results are reported in Table~\ref{tab:ablation}.

\begin{table}[h]
\centering
\caption{Branch Ablation Results}
\label{tab:ablation}
\begin{tabular}{lcc}
\hline
\textbf{Variant} & \textbf{Mean F1} & \textbf{Std F1} \\
\hline
CNN-only      & 0.738 & 0.045 \\
PARAFAC-only  & 0.691 & 0.009 \\
Fused         & \textbf{0.862} & 0.006 \\
\hline
\end{tabular}
\end{table}

The results demonstrate that the fused model outperforms both standalone branches by a substantial margin, achieving a mean F1 of 0.862 compared to 0.738 for CNN-only and 0.691 for PARAFAC-only. Notably, the fused model also exhibits the lowest variance across folds (std = 0.006), indicating that the combination of CNN spatial features and PARAFAC tensor decomposition features produces more stable and consistent predictions than either branch alone. The PARAFAC-only branch achieves competitive standalone performance relative to CNN-only (0.691 vs. 0.738), suggesting that CP tensor decomposition captures diagnostically relevant structural information that is complementary to, rather than redundant with, the spatial features extracted by the ResNet-18 backbone. These findings validate the architectural decision to fuse both branches, confirming that PARAFAC features contribute meaningfully to the model's overall classification performance.

\section{Conclusion/ Discussion}
\textbf{In this study, a classical ML pipeline with quantum-inspired mathematical structures, leveraging Random Forest Ensemble Methods and feature extractions through Tensor Decompositions}, is evaluated as a framework for multi-class brain MRI classification. Using five-fold nested stratified testing, the proposed model delivered reproducible, robust performance across higher and lower rank configurations on a rigorously cleaned, balanced dataset, demonstrating the effectiveness and stability of quantum-inspired classical approaches in extracting diagnostically relevant features from sparse, heterogeneous medical imaging data. Despite the strong results, the model produced the most frequent errors when distinguishing healthy control instances from mild pathologies, illustrating the intrinsic clinical complexity of such samples, and suggesting the need for further improvement in detecting subtle anatomical differences. The computational cost of tensor decompositions could constrain scalability in large imaging datasets, particularly in higher rank states.  Future work will focus on validating the approach on external, larger, and more diverse datasets; investigating the integration of quantum encoding schemes; and deploying the model on real quantum hardware. Additionally, the feature extraction pipeline can be further optimized through methods such as Tucker Decompositions or Manifold Learning. Through these practices, a hybrid quantum model could advance translational neuro-imaging and AI-driven diagnostics in clinical practice and biomedical tasks.

\vspace{12pt}
\color{red}


\begin{thebibliography}{00}
\bibitem{b1} World Health Organization. Over 1 in 3 people affected by neurological conditions, the leading cause of illness and disability worldwide.'' \textit{WHO News}, 14 March 2024. Available: \href{https://www.who.int/news/item/14-03-2024-over-1-in-3-people-affected-by-neurological-conditions--the-leading-cause-of-illness-and-disability-worldwide}{WHO News: Neurological conditions}

\bibitem{b2} Lustig, M., Donoho, D., \& Pauly, J. M. ``Sparse MRI: The application of compressed sensing for rapid MR imaging.'' \textit{Magnetic Resonance in Medicine}, 58(6):1182–1195, 2007. Available: \url{https://onlinelibrary.wiley.com/doi/10.1002/mrm.21391}

\bibitem{b3}
O.~N.~Jaspan, R.~Fleysher, and M.~L.~Lipton,
``Compressed sensing MRI: A review of the clinical literature,''
\textit{The British Journal of Radiology}, vol.~88, no.~1056, p.~20150487, Oct.~2015, doi: 10.1259/bjr.20150487. [Online]. Available: \url{https://www.ncbi.nlm.nih.gov/pmc/articles/PMC4984938/}


\bibitem{b4}
World Federation of Neurology.,
\textit{World Federation of Neurology}, Oct. 16, 2023. [Online]. Available: \url{https://wfneurology.org/activities/news-events/archived-news/2023-10-16-wcn}


\bibitem{b5} Steinmetz, J., et al. ``Global, regional, and national burden of disorders affecting the nervous system, 1990-2021: a systematic analysis for the Global Burden of Disease Study 2021'' \textit{The Lancet Neurology}, 23(4):344–381, 2024. Available: \url{https://pubmed.ncbi.nlm.nih.gov/38493795/}

\bibitem{b6} D. Konar et al., "3D Quantum-Inspired Self-Supervised Tensor Network 
for Volumetric Segmentation of Medical Images," \textit{IEEE}, 2024. [Online]. Available: \url{https://ieeexplore.ieee.org/document/10038494}

\bibitem{b7}  R. Baraniuk and P. Steeghs, "Compressive Radar Imaging," \textit{IEEE}, 2007. [Online]. 
Available:\url{https://ieeexplore.ieee.org/document/4250297}

\bibitem{b8} Selvan, R. \& Dam, E. B. ``Tensor Networks for Medical Image Classification.'' \textit{Nature Scientific Reports}, 13:30258, 2023. Available: \url{https://doi.org/10.48550/arXiv.2004.10076}

\bibitem{b9}
Mykhailo Klymenko, Thong Hoang, Xiwei Xu, Zhenchang Xing, Muhammad Usman, Qinghua Lu, and Liming Zhu.
``Architectural Patterns for Designing Quantum Artificial Intelligence Systems.'' \textit{arXiv preprint arXiv:2411.10487}, 2024.
\url{https://arxiv.org/abs/2411.10487}

\bibitem{b10}
A.~C.-Y. Yang, M.~Kretzler, S.~Sudarski, V.~Gulani, and N.~Seiberlich,
``Sparse Reconstruction Techniques in MRI: Methods, Applications, and Challenges to Clinical Adoption,''
\textit{Investigative Radiology}, vol.~51, no.~6, pp.~349--364, Jun. 2016. [Online]. Available: \url{https://pmc.ncbi.nlm.nih.gov/articles/PMC4948115/}

\bibitem{b14} H.-M. Rieser, F. Köster, and A. P. Raulf, "Tensor networks for quantum machine learning," \textit{arXiv preprint arXiv: 2303.11735}, Mar. 2023. Available: \url{https://doi.org/10.1098/rspa.2023.0218}

\bibitem{b15} Faccio, D., "The future of quantum technologies for brain imaging," Frontiers in Physics, vol. 12, 2024. [Online]. Available: \url{https://pmc.ncbi.nlm.nih.gov/articles/PMC11515994/}

\bibitem{b16} M. F. Shahriyar and G. Tanbhir, "Advancements and Challenges in Quantum Machine Learning for Medical Image Classification: A Comprehensive Review," arXiv preprint arXiv:2504.13910, 2024. [Online]. Available: \url{https://arxiv.org/abs/2504.13910}

\bibitem{17} M. Wang, Y. Pan, Z. Xu, G. Li, X. Yang, D. Mandic, A. Cichocki, "Tensor Networks Meet Neural Networks: A Survey and Future Perspectives," arXiv preprint arXiv:2301.11711, 2023. [Online]. Available: \url{https://arxiv.org/html/2302.09019v3}

\bibitem{18}
M.~Kidoh, K.~Shinoda, M.~Kitajima, K.~Isogawa, M.~Nambu, H.~Uetani, K.~Morita, T.~Nakaura, M.~Tateishi, Y.~Yamashita, and Y.~Yamashita,
``Deep Learning Based Noise Reduction for Brain MR Imaging: Tests on Phantoms and Healthy Volunteers,''
\textit{Magn. Reson. Med. Sci.}, vol.~20, no.~1, pp.~26--37, 2021, doi: 10.2463/mrms.mp.2020-0111. [Online]. Available: \url{https://www.ncbi.nlm.nih.gov/pmc/articles/PMC7553817/}

\bibitem{19}
S.~Krishnapriya and Y.~Karuna, ``Pre-trained deep learning models for brain MRI image classification,'' \emph{Frontiers in Human Neuroscience}, vol.~17, p. 1150120, 2023. [Online]. Available: \url{https://www.frontiersin.org/articles/10.3389/fnhum.2023.1150120/full}

\bibitem{b20}
A. Dahal, S. A. Murad, and N. Rahimi, ``Efficiency bottlenecks of convolutional Kolmogorov-Arnold networks: A comprehensive scrutiny with ImageNet, AlexNet, LeNet and tabular classification,'' \emph{arXiv preprint arXiv:2501.15757}, 2025. [Online]. Available: \url{https://arxiv.org/pdf/2501.15757.pdf}


\end{thebibliography}
\end{document}